\documentclass[showpacs,preprintnumbers,amsmath,amssymb]{revtex4}
\usepackage{graphicx}
\usepackage{dcolumn}
\usepackage{bm}

\begin{document}

\title{Control of energy distribution of the proton beam with an oblique
incidence of the laser pulse}

\author{Toshimasa~Morita, Sergei~V.~Bulanov, Timur~Zh.~Esirkepov,
James~Koga, Mitsuru~Yamagiwa}
\affiliation{Advanced Photon Reserch Center,
Japan Atomic Energy Agency, 8-1 Umemidai, Kizugawa,
Kyoto 619-0215, Japan}


\begin{abstract}
We investigate proton acceleration by a laser pulse obliquely incident on a
double layer target via 3D PIC simulations.
It is found that the proton beam energy spread changes by the laser irradiation
position and it reaches a minimum at certain position.
This provides a way to control the proton energy spectrum.
We show that by appropriately adjusting the size and position of the second
proton layer that high energy protons with much smaller energy spread can be
obtained.
\end{abstract}


 
\maketitle

\section{INTRODUCTION}
Compact laser technology has advanced remarkably in recent years, such that
the laser power and intensity have improved exponentially.
Currently, laser powers have reached the petawatt range, and
intensities exceeding $10^{22}W/cm^{2}$ have been achieved \cite{VYA}.
Assuming that this great progress in this compact laser technology continues
devices which were thought to be difficult to achieve may become possible
and be much smaller than possible with conventional technology.

Charged particle acceleration is the one of the important examples of
applying this compact laser technology.
The method of laser acceleration of charged particles by using laser light
is very attractive, since the acceleration rate is markedly higher and the
facility size can be substantially smaller than that of standard accelerators.
Laser driven fast ions could be applicable to hadron therapy \cite{SBK},
fast ignition of thermonuclear fusion \cite{ROT}, production of PET
sources \cite{SPN}, conversion of radioactive waste \cite{KWD}, proton
imaging of ultrafast processes in laser plasmas \cite{PrIm}, a laser-driven
heavy ion collider \cite{ESI1}, and a proton dump facility for neutrino
oscillation studies \cite{SVB}.
Numerous applications require high quality proton (ion) beams, i.e. beams
with sufficiently small energy spread $\Delta \mathcal{E}/\mathcal{E}<<1$.
For example, for hadron therapy it is highly desirable to have a proton beam
with $\Delta \mathcal{E}/\mathcal{E}\leq 2\%$ in order to provide the
conditions for a high irradiation dose being delivered to the tumor while
sparing neighboring tissue \cite{KhM}. In the concept of fast ignition with
laser-accelerated ions \cite{ROT} an analysis has shown that the ignition of a
thermonuclear target with a quasi-thermal beam of fast protons requires several
times larger laser energy than a beam of quasi-monoenergetic protons \cite{ATZ}.
Therefore, a laser generated beam of quasi-monoenergetic protons
is highly desired.
Similarly, in the case of the ion injector \cite{INJ}, a
high-quality beam is needed in order to inject the charged particles into
the optimal accelerating phase.

As suggested in Ref. \cite{DL},
such an ion beam can be obtained using a double-layer target
consisting of high-Z atoms and a thin coating of low-Z atoms.
Extensive computer simulations of this target were performed in Refs.
\cite{ESI} and \cite{EYT}.
The proof of principal for the high-quality ion beam generation with the
irradiation of the double-layer targets was been done in Ref. \cite{DLExp}.
It has also been reported (Ref. \cite{MEBKY}) that a higher energy proton beam
is generated for oblique incidence than for normal incidence on a double layer
target.
However, it is also shown that the energy spread of the
proton beam is larger for oblique incidence than for normal incidence.
As mentioned above, the wider energy spread of the proton beam is undesirable
in terms of applications such as proton cancer treatment.
Therefore, it is important to study how the energy spread can be reduced for
oblique incidence. In this paper, we use a particle-in-cell(PIC) code to
investigate the control of the energy distribution of the proton beam for
oblique laser incidence on a double layer target.

\section{SIMULATION MODEL}
We study the dependence of the ion beam energy and quality on
the laser irradiation point of the target with oblique incidence,
where the incidence angle is 30 degrees in all the simulations.
The reason for using 30 degrees is that it has been reported that protons
obtain the maximum energy at 30 degrees in Ref. \cite{MEBKY}, \cite{MEBKY2}.
We use an idealized model, in which a Gaussian p-polarized laser pulse is
incident on a double-layer target of collisionless plasma.
The simulations are performed with a three-dimensional massively parallel
electromagnetic code, based on the PIC method \cite{CBL}.

In the present simulations,
the total number of quasi-particles equals $8\times 10^{7}$.
The size of the simulation box is ($L_X,L_Y,L_Z$)
$100\lambda \times 54\lambda \times 27\lambda $,
where $\lambda$ is the laser wavelength.
Here, we define the laser propagation direction as being along the $X$ axis,
the electric field of the laser pulse along the $Y$ axis, and
the magnetic field of the laser pulse along the $Z$ axis.
The oblique incidence of the laser pulse is realized by tilting the target
around the $Z-$axis, while the laser pulse propagates along the $X-$axis.
The reason for using $L_Y$ two times larger than $L_Z$ is that the
laser and proton propagation directions are different.
The protons move toward the upper boundary in the $Y$ direction.
We adopted $L_Y$ to be large so that the boundary would not
influence the generated protons.
The number of grid cells is equal to $2800\times1536\times 768$ \
along the $X$, $Y$, and $Z$ axes respectively.
The boundary conditions for the particles and for the fields are
periodic in the transverse ($Y$,$Z$) directions and absorbing at the
boundaries of the computation box along the $X$ axis. Here the laser
wavelength determines the transformation from dimensionless to dimensional
quantities and vice versa.

In order to more easily understand the results, we present the simulation
results in a frame rotated 30 degrees with respect to the $Z-$axis.
We define the direction perpendicular to the target surface as the positive
$x-$axis and the direction parallel to the target as the $y-$axis,
which are the $X,Y-$axes rotated by 30 degrees about the $Z-$axis.
The $Z-$axis and $z-$axis are the same.

Below, the dimensional quantities are given for
$\lambda = 0.8\mu $m; the spatial coordinates are normalized by $\lambda $
and the time is measured in terms of the laser period, $2\pi/\omega$.

Both layers of the double-layer target are shaped as disks.
The first, gold with $Z_{i}=+2$ and $m_{i}/m_{e}Z_{i}=195.4\times 1836/2$,
layer has a diameter of $10\lambda $ and thickness of
$0.5\lambda $. The second, hydrogen with $m_{p}/m_{e}=1836$,
layer is narrower and thinner than the gold layer;
The diameter of the second layer and its position with respect to the
first laser are changed in different simulations as specified later
where the thickness is $0.03\lambda $ in all the simulations.
The electron density inside the gold layer is
$n_{e}=1.6\times 10^{22}$cm$^{-3}$ and inside the proton
layer it is $n_{e}=5\times 10^{20}$cm$^{-3}$.
The Gaussian laser pulse with the dimensionless amplitude
$a=eE_{0}/m_{e}\omega c=30$, which corresponds to the laser peak intensity
$2\times 10^{21}W/cm^{2}$, is $8\lambda $ long in the propagation direction
and is focused to a spot with size $6\lambda $ (FWHM).

\section{RESULTS OF SIMULATIONS}
\begin{figure}
\includegraphics[width=8.6cm]{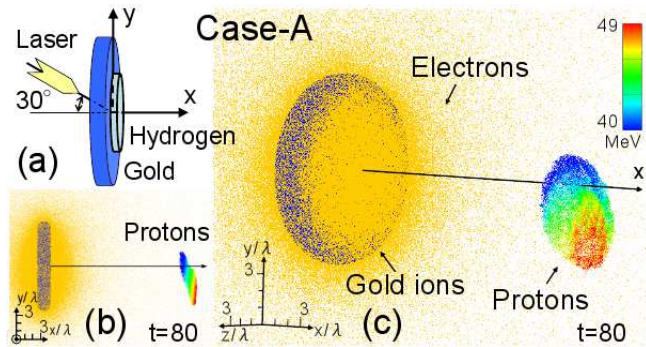}
\caption{\label{fig:A} Case-A: Oblique incidence configuration in the laser
pulse interaction with a double layer target (a) and
the distribution of gold ions (blue), electrons (yellow), and protons at t=80.
The protons are classified by the energy (color) (b),(c).}
\end{figure}

First, we show a case (case-A) where the diameter of the second
(proton) layer is 5$\lambda $ (half the diameter of the first layer)
and it is placed at the center of first layer (Fig.\ref{fig:A}(a)).
In this case, the laser irradiates the center of the target.
This case has the same conditions for the target and laser pulse with
an incidence angle of 30 degrees as Ref. \cite{MEBKY}.
Figure \ref{fig:A}(b,c) shows the distribution of the generated protons.
Each proton is classified depending on the amount of the energy.
High energy protons are red, low energy protons are blue,
and energies in between are indicated by the color bar in Fig. \ref{fig:A}(c).
The red area shows the high energy part ($\geq$48.8MeV) and
the blue area shows the low energy part ($\leq$39.5MeV).
We see that the generated protons are distributed in different areas
based on each energy level.
The energy of the generated protons is a function of coordinates,
and this suggests that we are able to selectively extract high energy
protons by a simple method. 

\begin{figure}
 \includegraphics[clip, width=7.0cm]{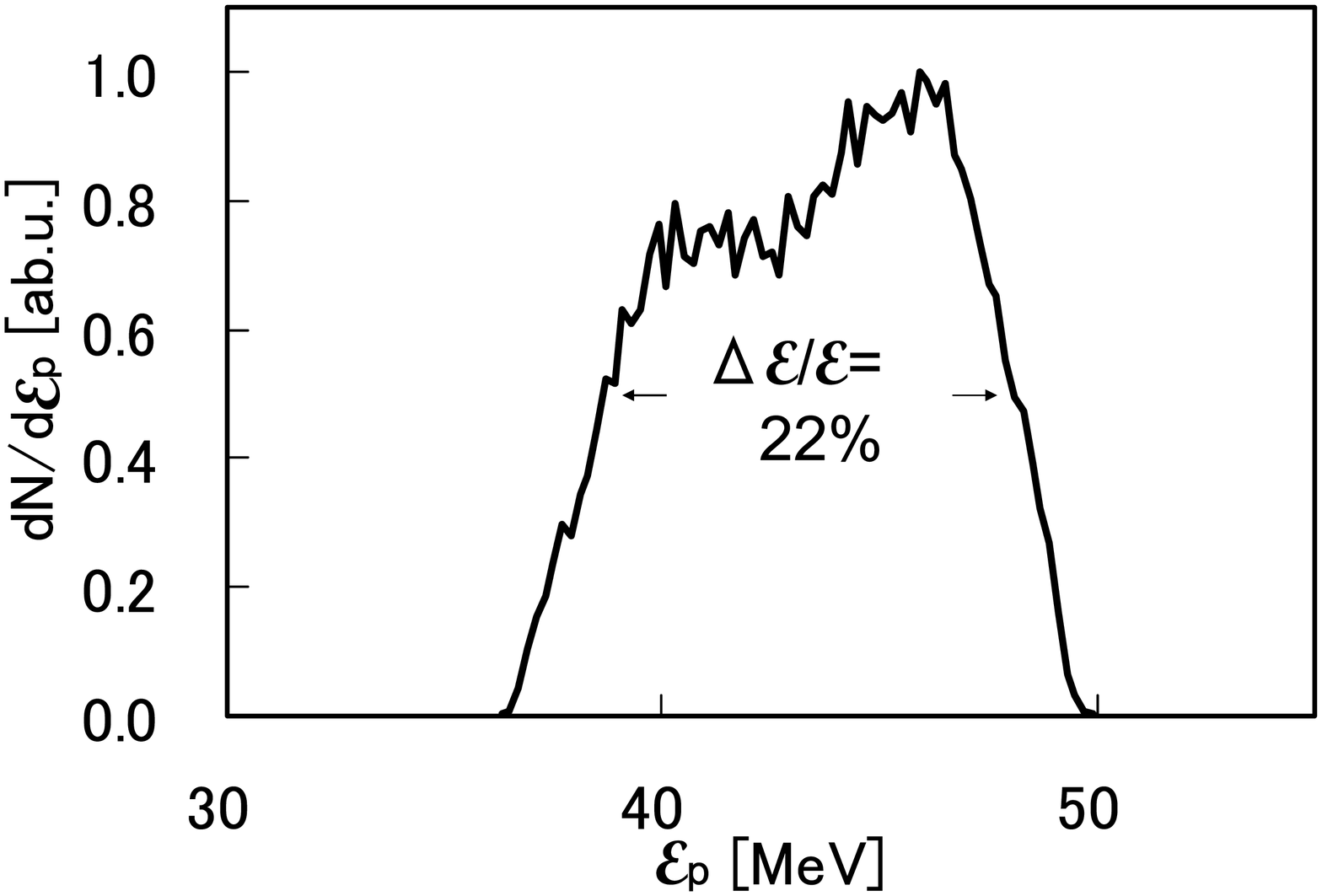}
\caption{\label{fig:B} Proton energy spectra, normalized by the maximum,
at t=80 of case-A.
}
\end{figure}

Figure \ref{fig:B} shows the energy spectrum of the generated protons
at $t=80$ of case-A.
The proton energy is about $20MeV$ for the normal incidence case shown in
Ref. \cite{MEBKY} so the proton energy for the oblique incidence case is
two times higher than the normal incidence case.
On the other hand,
the energy spread of the generated protons is 9MeV in the FWHM, and the energy
spread ratio in the FWHM is about 22$\%$. The energy spread in the normal
incidence case shown in Ref. \cite{MEBKY} is about 10$\%$,
so the energy spread has increased by about a factor of two in the oblique
incidence case.

\begin{figure}
\includegraphics[width=8.6cm]{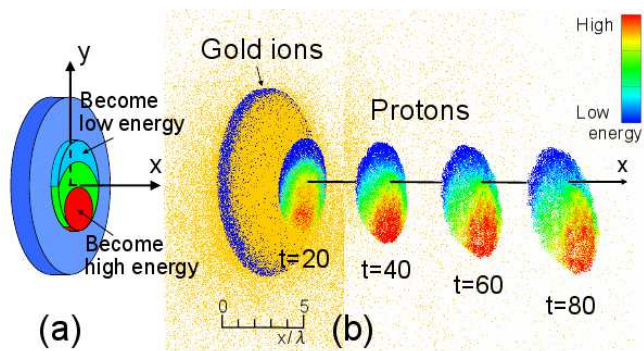}
\caption{\label{fig:C}
Energy distribution of protons in the initial proton layer.
The red color in the proton layer shows the part becoming high energy protons,
and blue color shows the part becoming low energy protons (a),
and the proton energy distribution at different simulation times
(t=20,40,60,80), the red color shows the part at relatively high energy
and blue color shows the  part at relatively low energy at each time (b).
}
\end{figure}

Figure \ref{fig:C}(b) shows the energy distribution of the generated proton
bunch at different times during the simulation.
We can see that the proton bunch has rotated a few degrees clockwise about
the $z-$axis initially (t=20).
This is because the acceleration begins for protons at $+y$ positions earlier
than $-y$ positions since the laser reaches the $+y$ positions earlier than
the $-y$ positions (Ref. \cite{MEBKY}).
As time advances, the proton bunch becomes anti-clockwise rotated  with respect
to the $z-$axis.
This is because the protons initially at $-y$ positions reach a high
speed since they are accelerated more strongly than protons initially at $+y$
positions (shown later). 
In addition we can see that overall the proton bunch has moved slightly
downward in the $-y$ direction (Ref. \cite{MEBKY}).
The relative energy of the generated protons at each time are indicated by
colors  with the high energy indicated by red and low energy indicated by blue.
The protons at the position below the laser pulse center in the initial proton
layer become more energetic (Fig. \ref{fig:C}(a)).
In our simulations, we used a very thin and low density proton layer,
so the generated proton bunch keeps the initial disc shape and size.
If we use a thicker and higher density proton layer, however,
the generated proton bunch could expand from the initial shape.
In this case, the point where the high energy protons are obtained is shifted
to larger $-y$ values than the case shown in Fig. \ref{fig:C}(b) due to
the effect of this expansion.
In experiments using a thin polyimide foil, it is reported that the higher
energy protons are observed at an angle shifted away from the target normal
towards the direction of obliquely incident laser pulse (see Ref. \cite{YOG}).
In Ref. \cite{YOG}, they reported a larger deflection angle away from the
target normal than our simulation,
it is thought this is due to expansion of the generated proton bunch.
As we have discussed in this paper,
protons at the position downward ($-y$ direction) from the laser pulse center
become more energetic than those at the center.
If the protons at the laser center position had become the most energetic,
they would continue to be located at the center of the generated proton bunch
after expansion, so a big shift in angle away from the target normal of the
high energy part would not have been observed.

The spread of the proton energy spectrum (Fig. \ref{fig:B}) corresponds to
the energy distribution shown in Fig. \ref{fig:A}(c).
There is a spatial correspondence to the energy spread between the higher
energy and lower energy protons, therefore, it is possible to reduce the energy
spread by appropriately adjusting the spatial distribution.
According to this idea, we will show the simulation results of several cases
where the energy spread of the protons has been decreased. 

\begin{figure}
\includegraphics[width=8.6cm]{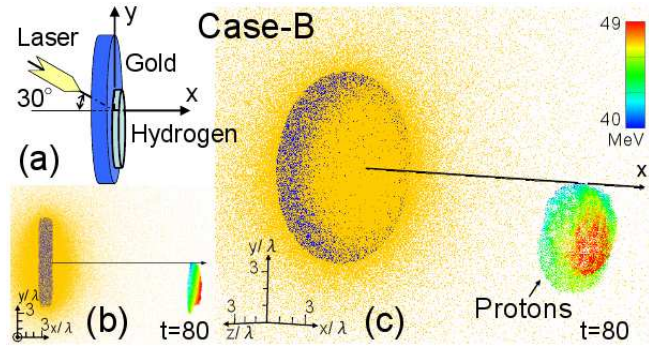}
\caption{\label{fig:D}
Case-B: Configuration of the target and the laser pulse.
The proton layer is placed downward in the $-y$ direction with the center at
$0.5r$ where $r$ is the radius of the proton layer (a) and
the distribution of gold ions (blue), electrons (yellow), and protons at t=80.
The protons are classified by energy (color) (b),(c).
}
\end{figure}

\begin{figure}
\includegraphics[clip,width=8.6cm]{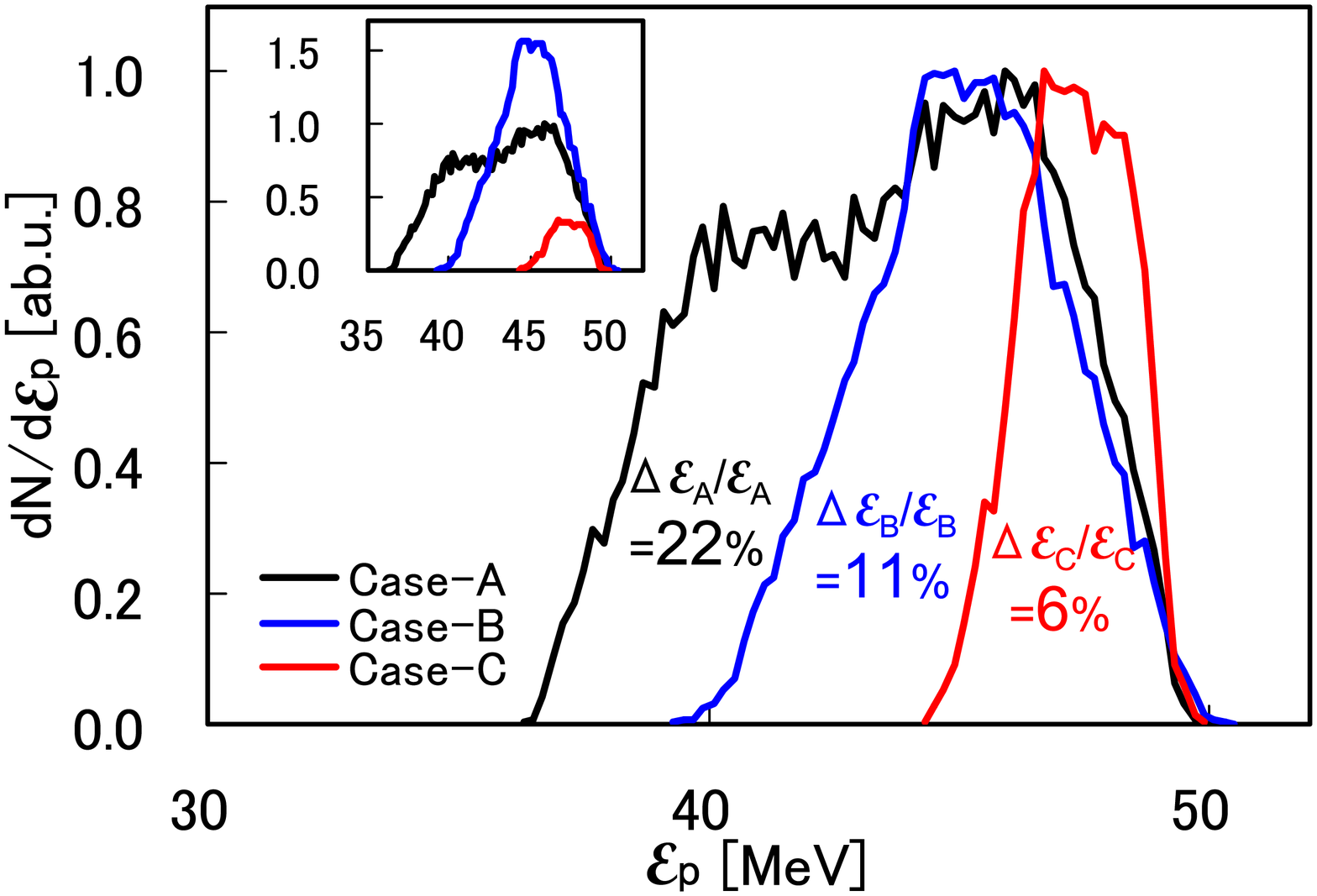}
\caption{\label{fig:F}
Proton energy spectra, normalized by the maximum of each case, at t=80.
In the inset: Proton energy spectra which are normalized by using the number
of protons of case-A.
}
\end{figure}

We consider several cases of the initial proton distributions.
Case-B: change the position of the proton layer and the size of proton
layer is not changed.
Case-C: change the position and size of the proton layer.
When the size of proton layer is not changed (case-B),
we can reduce the energy spread by shifting the proton layer so that the part
corresponding to the red area in Fig. \ref{fig:A}(c) is located at the center
of the proton bunch.
Because the energy of the generated protons decreases with distance
from the position where highest energy protons occur,
we can suppress the decrease in the proton energies at the edge of the bunch,
if we shorten the distance from where the highest energy protons are generated
to the edge of the proton bunch by locating the highest energy part at the
center of the proton layer.
In case-A, the high energy protons are distributed in the lower part ($-y$) of
the initial proton layer (Fig. \ref{fig:A}(c)), so moving the initial proton
layer downward ($-y$ direction) a little could reduce the energy spread.
Figure \ref{fig:D} shows the results for case-B where the initial proton
layer is placed downward in the $-y$ direction with the center at $0.5r$ where
$r$ is the radius of the proton layer.
The high energy proton area is distributed around the center of the
proton bunch as shown in Fig. \ref{fig:D}(b,c).
The energy of the protons at the edge of the proton bunch is higher than case-A.
The proton energy spectrum is shown in Fig. \ref{fig:F} (case-B).
The energy spread in the FWHM is about $11\%$,
which is about half of that of case-A.

\begin{figure}
\includegraphics[width=8.6cm]{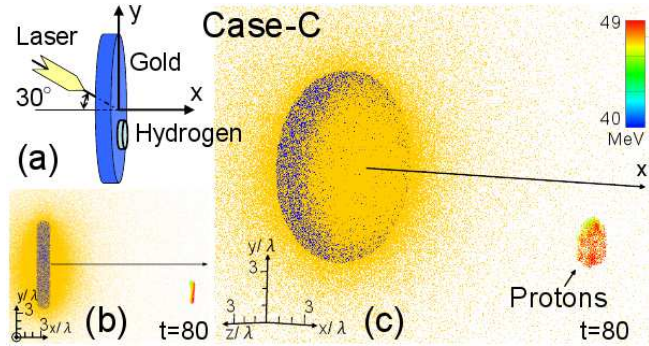}
\caption{\label{fig:E}
Case-C: Configuration of the target and the laser pulse.
The diameter of the proton layer is reduced and placed downward in the $-y$
direction (a) and the distribution of gold ions (blue), electrons (yellow),
and protons at t=80.
The protons are classified by energy (color) (b),(c).
}
\end{figure}

Next,
we show the case when the initial proton layer size is also changed (case-C).
We can reduce the energy spread of the generated protons by placing an initial
proton layer only around the part from where the higher energy protons colored
in red occur in Fig. \ref{fig:A}(c).
Figure \ref{fig:E} shows the result for which the diameter of the
initial proton layer is reduced to $1/3$ and is placed at the
position shifted downward ($-y$ direction) from the laser irradiation center.
We can see that only high energy protons are generated as shown
in Fig. \ref{fig:E}(b,c).
The energy spectrum of the generated protons in this case is shown
in Fig. \ref{fig:F} (case-C).
In this case, the energy spread is only about $6\%$,
which is less than one-third of case-A.
We note that, setting a screen which has a pin-hole behind the target in
case-A, might give a similar result, and is equal to selecting only the
high energy protons from case-A, so in the energy spectrum,
it is only the high energy part of case-A.
Here, the key to obtain the higher energy protons is,
as for the pin-hole position, to put it in a place shifted slightly from
the laser irradiation position normal to the target surface towards the laser
propagation direction.

\begin{figure}
\includegraphics[width=8.6cm]{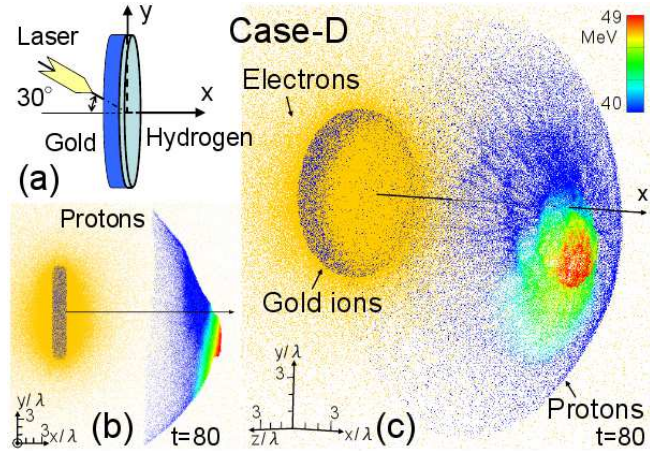}
\caption{\label{fig:G}
Case-D: Configuration of the target and the laser pulse.
The diameter of the proton layer is the same as the first layer (a) and
the distribution of gold ions (blue), electrons (yellow), and protons at t=80.
The protons are classified by energy (color) (b),(c).
}
\end{figure}

Figure \ref{fig:F} shows the spectra of the generated protons for cases-A,B,C,
where they are normalized to the maximum value of each case.
In case-C, a small proton layer is used compared with case-A,B,
so the number of total protons is smaller compared with the other cases.
The inset in the figure shows the energy distributions of the protons for
case-A,B,C where each distribution has been normalized by the maximum value
of the distribution for case-A.
We can compare the numbers of protons in case-A, B, and C
at each energy level.
We can get high energy and high quality protons in case-C.
However, from the point of view of the number of generated protons,
Fig. \ref{fig:F}(inset) shows that we can obtain a lot of mono-energetic protons
in case-B.

Figure \ref{fig:G} shows the results of the case when the proton layer has the
same diameter as the gold layer (case-D).
We can reexamine the above cases A, B, C by considering this case.
We can get the case-A result by cutting out the generated protons around the
centerline of the target layer with some radius,
and the case-B result by cutting out the protons above some radius centered
about the high energy part of the proton distribution.
The case-C result is obtained by making the cut off radius smaller so that
the only remaining area is the high energy area (red color area).
Here, the point on which we have paid attention is that the high energy
protons appear at a position a little bit below in the $-y$ direction
from the centerline of the target (Fig. \ref{fig:G}).
This downward ($-y$) shift of the high energy proton part is the cause of the
energy spread which is larger in the oblique incidence case in
Ref. \cite{MEBKY}.

\begin{figure}
 \includegraphics[clip, width=7.0cm]{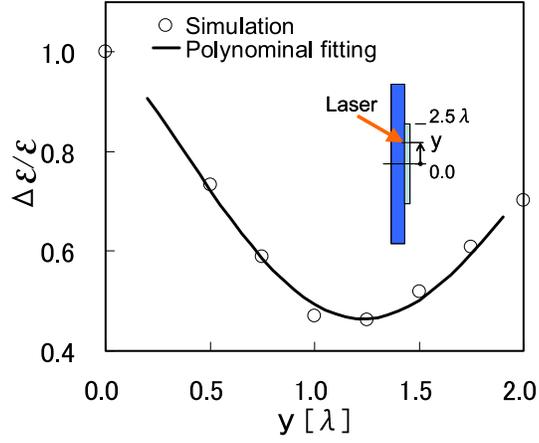}
\caption{\label{fig:H}
Proton energy spread, normalized by the maximum,
vs the laser pulse irradiation point.
}
\end{figure}

Figure \ref{fig:H} shows the proton energy spread at different laser pulse
irradiation points.
In this case, the diameter of the proton layer is 5$\lambda $ 
and placed at the center of the first layer. This is the same as case-A,
but with different irradiation points.
We changed the position of the proton layer in case-B, however,
here we changed the laser irradiation point. 
When the first layer is very large,
changing the proton layer position is equivalent to
changing the irradiation point of the laser.
Therefore, here we see the changing of the energy spread when the laser
irradiation position is continuously changed.
We can see that the energy spread of the protons decreases as the laser
irradiation position is moved from the center of the target towards $+y$.
The minimum spread occurs around $y=1.25\lambda $ which is $0.5r$ from
the proton layer center where $r$ is the radius of the proton layer.

Here, we investigate the reason why the high energy proton part shifts a
little downward ($-y$) from the laser pulse center.
Having set the direction normal to the target surface to be $x$.
We consider the process by which a proton is accelerated in a uniform
electric field in the $x$ direction, $E_x$.
Here, we assume the value of the electric field in the $y$ direction varies.
We consider a proton which is
accelerated from the upper point of the proton layer ($+y$)
and a proton which is accelerated from the lower point ($-y$).
It is assumed that the laser reaches the upper part of the proton layer at
time $t=0$, and then reaches the lower point at time $t=t_0$
which is later than for the upper point.
Therefore, the upper point proton begins accelerating at time $t=0$,
and a lower point proton begins accelerating at time $t=t_0$.
In this situation, the velocity of the proton at time $t$ becomes
$v_{u}=q\int_0^tE_{xu}(\tau)d\tau/m$ at the upper position, and
$v_{l}=q\int_{t_0}^tE_{xl}(\tau)d\tau/m$ at the lower position,
where, $q$ is the charge of the proton, $m$ is the proton mass,
$E_{xu}$ is the upper point electric field in the $x$ direction,
and $E_{xl}$ is the lower point field.
For a $50MeV$ energy proton which is the maximum proton energy in case-A,
$\gamma=1/\sqrt{(1-v^2/c^2)}=1.05$,
so we estimate $\mathcal{E}_l/\mathcal{E}_u=v_l^2/v_u^2$
where we have made the approximation $\gamma\approx1.0$,
$\mathcal{E}_l$ is the lower position proton energy and
$\mathcal{E}_u$ is the upper position one.
The condition of $\Delta \mathcal{E}/\mathcal{E}\leq 2\%$ is equivalent to
$\mathcal{E}_l/\mathcal{E}_u\leq 1.02$, and we obtain the condition
$\int_{t_0}^tE_{xl}(\tau)d\tau/\int_0^tE_{xu}(\tau)d\tau=v_l/v_u=%
\sqrt{\mathcal{E}_l/\mathcal{E}_u}\leq\sqrt{1.02}\approx1.01$.
Therefore, if we want to get protons which satisfy the condition
$\Delta \mathcal{E}/\mathcal{E}\leq 2\%$,
it is necessary to satisfy the condition that the difference between
$\int_0^tE_{xu}(\tau)d\tau$ and $\int_{t_0}^tE_{xl}(\tau)d\tau$
should be less than a few percent.

\begin{figure}
\includegraphics[clip,width=5.0cm]{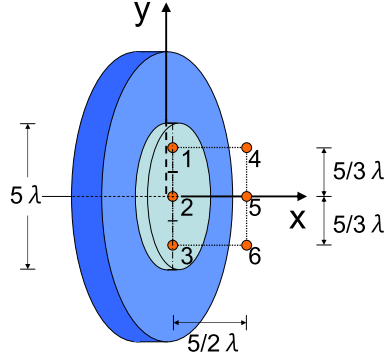}
\caption{\label{fig:I}
The sample points for $E_x^{fil.}$ and $\int E_x(t)dt$ described in the text.
}
\end{figure}

\begin{figure*}
  \begin{tabular}{cc}
  \end{tabular}
  \begin{center}
    \begin{minipage}{0.40\textwidth}
      \includegraphics[clip,width=\textwidth]{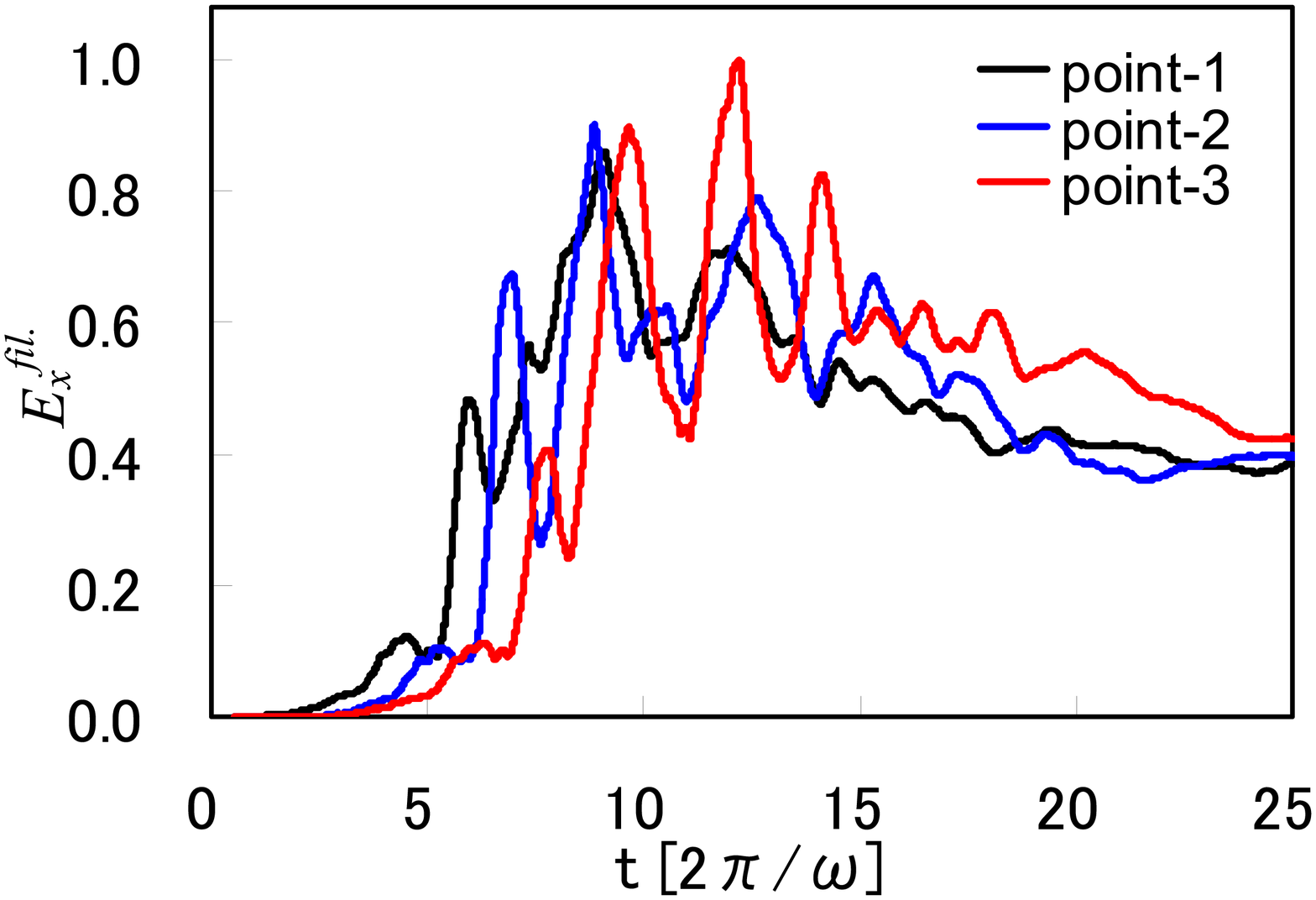}
(a) $E_x^{fil.}$ vs time at points-1,2,3.
    \end{minipage}
    \begin{minipage}{0.40\textwidth}
      \includegraphics[clip,width=\textwidth]{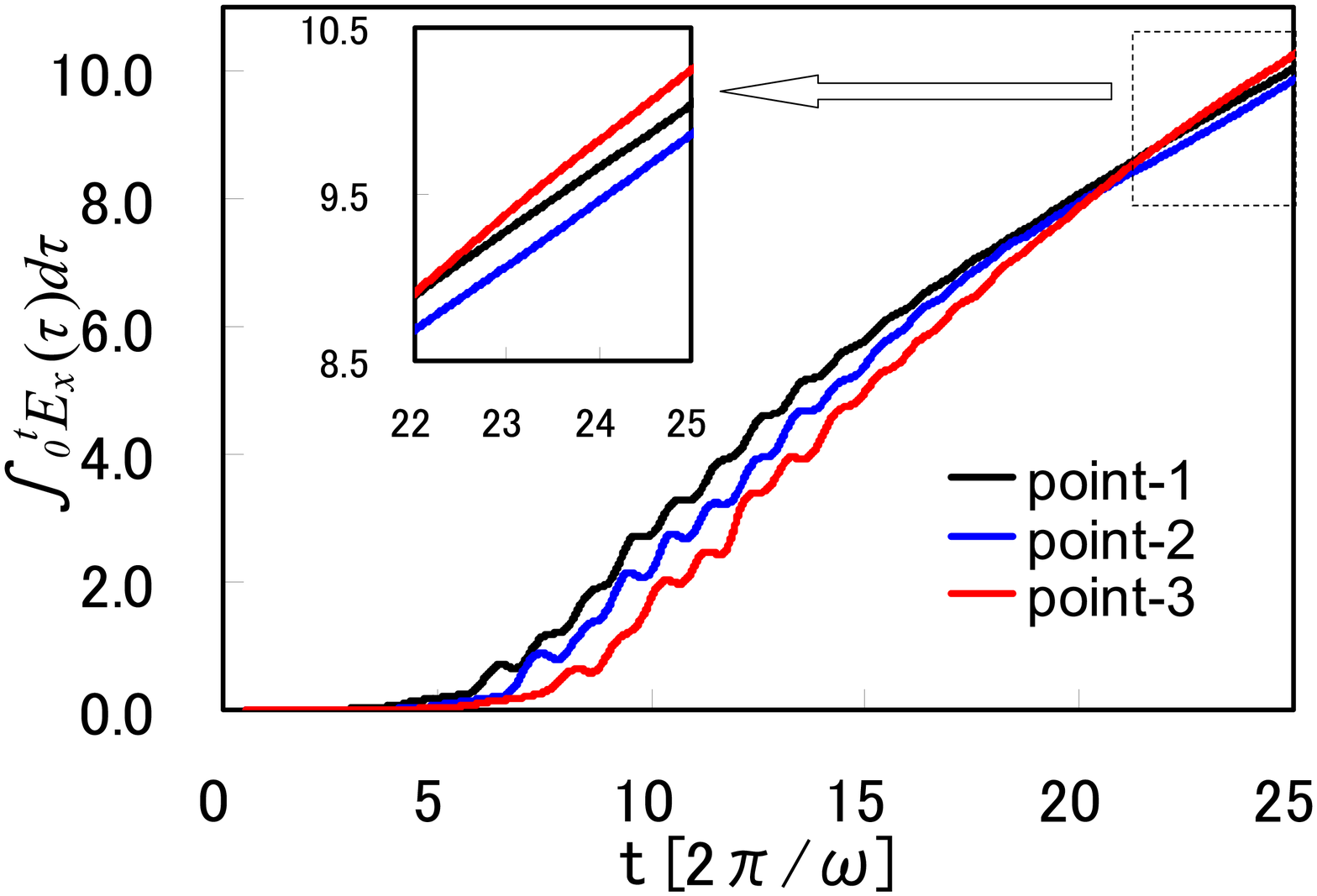}
(b) $\int E_x(t)dt$ vs time at points-1,2,3.
    \end{minipage}
    \begin{minipage}{0.40\textwidth}
      \includegraphics[clip,width=\textwidth]{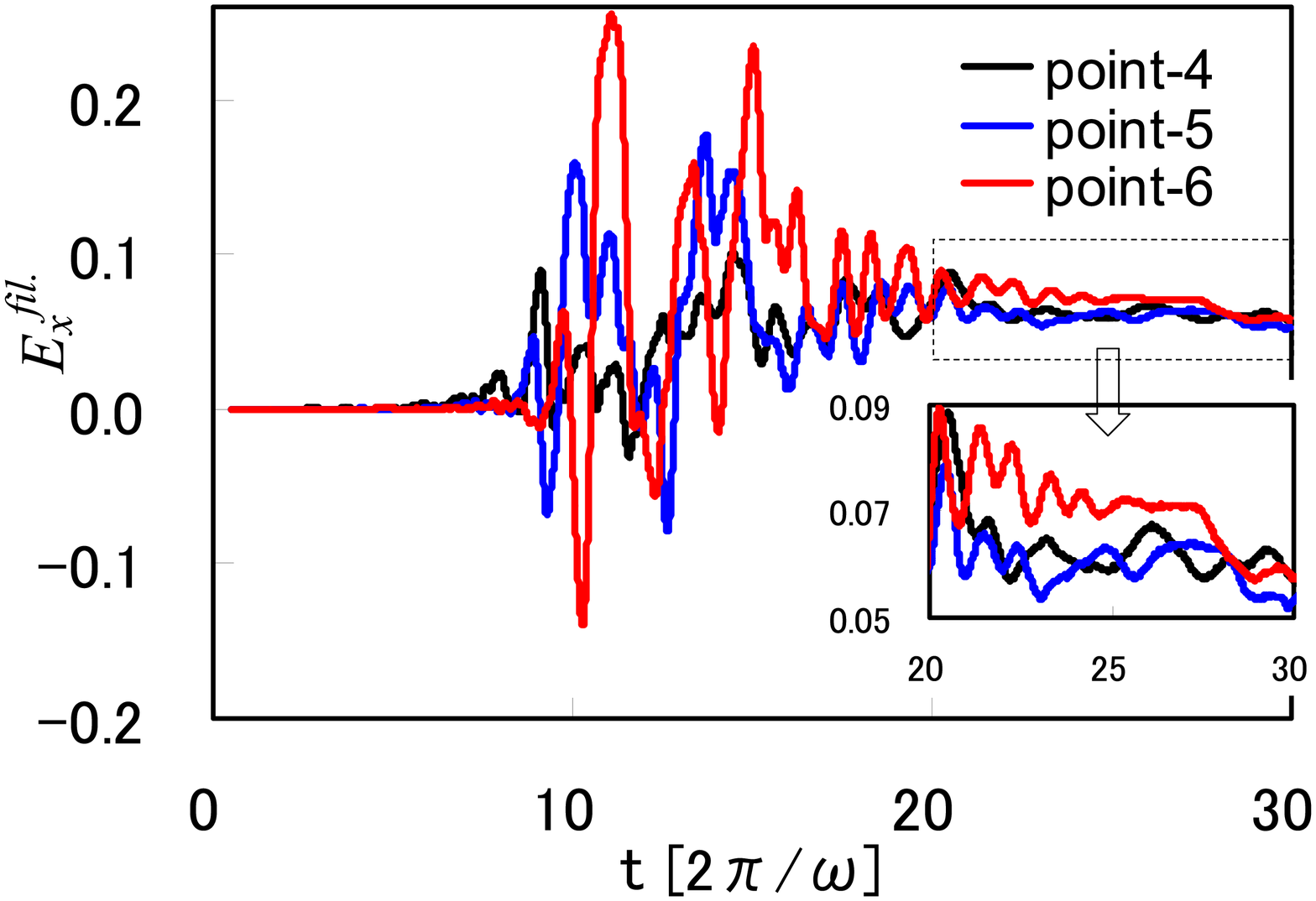}
(c) $E_x^{fil.}$ vs time at points-4,5,6.
    \end{minipage}
    \begin{minipage}{0.40\textwidth}
      \includegraphics[clip,width=\textwidth]{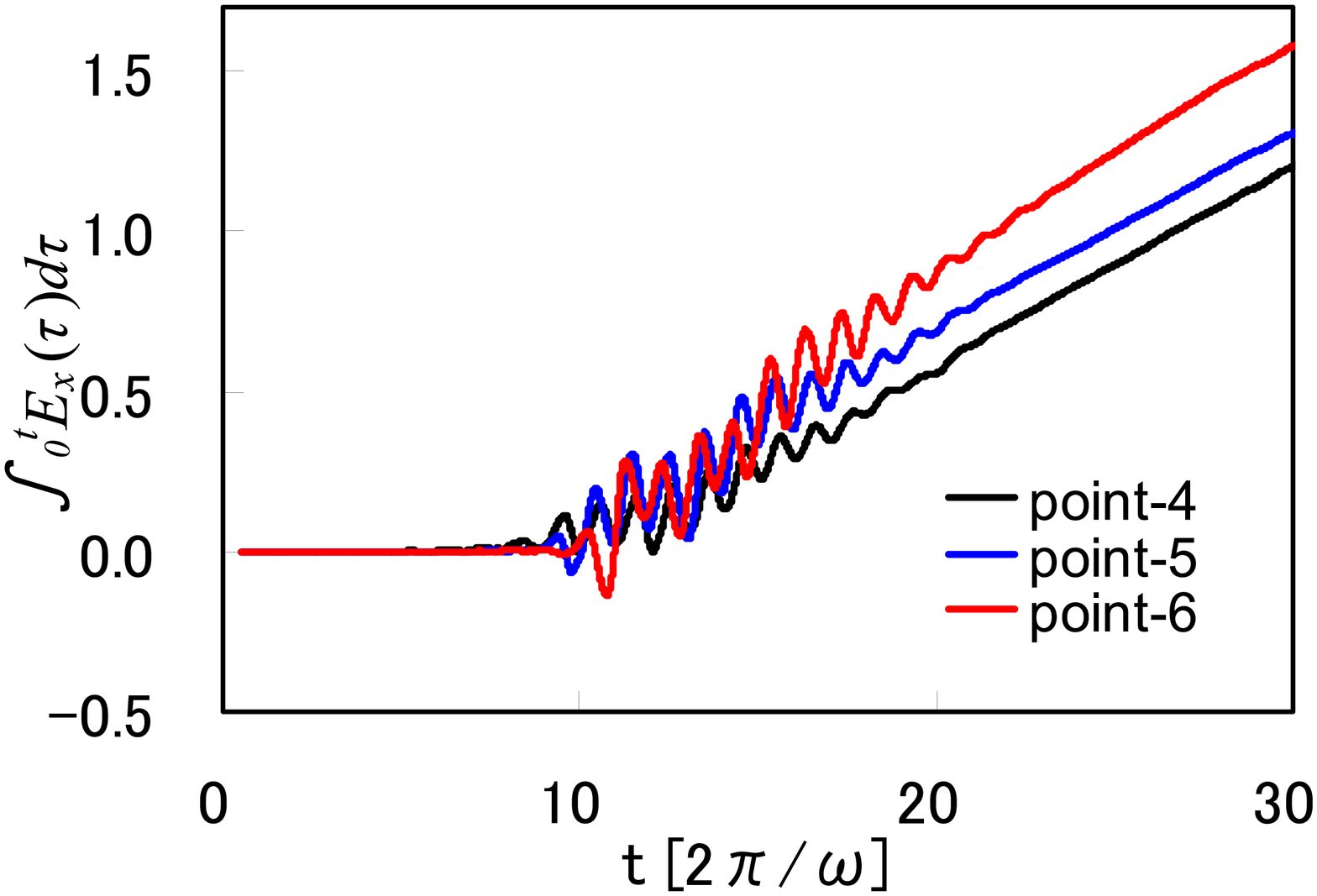}
(d) $\int E_x(t)dt$ vs time at points-4,5,6.
    \end{minipage}
  \caption{\label{fig:J}
The time variation of $E_x^{fil.}$ and $\int E_{x}(t)dt$,
described in the text, around the target for case-A.
}
  \end{center}
\end{figure*}

\begin{figure}
\includegraphics[width=8.6cm]{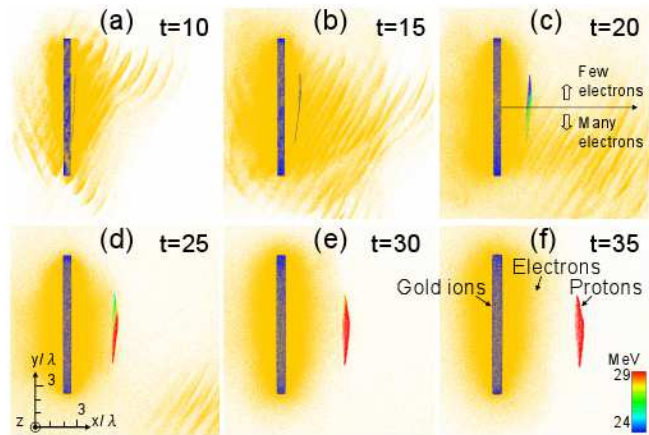}
\caption{\label{fig:K}
Particle distribution at early simulation times for case-A.
The blue color indicates the gold ions, the yellow indicates the electrons,
and the protons are colored by energy level.
}
\end{figure}

Figure \ref{fig:J}(a,c) shows the time change of the electric field in the $x$
direction, $E_{x}$, at each position (see Fig. \ref{fig:I}) near the target,
for case-A. The protons are accelerated by this $E_{x}$.
Figure \ref{fig:J}(a,c) show results of filtering ($E_x^{fil.}$) the vibration
of the electric field of the laser pulse,
by averaging the electric field over the laser frequency,
which is normalized by maximum value of $E_x^{fil.}$ in Fig. \ref{fig:J}(a).
The laser pulse comes from the upper left side (see Fig. \ref{fig:A}(a)), so
the laser pulse reaches the upper position of the target earlier (point-1),
and hence the electric field starts first at the upper part of the target
(point-1).
It is shown that point-3 which is at a lower position on the initial proton
layer experiences on average the largest $E_{x}$ in the time 13 to 25,
and the smallest field occurs at point-1.
The time until when the protons remain around the target surface is about
t=20 (see Fig. \ref{fig:C},\ref{fig:K}(c)),
so the lower point protons experience a larger electric field,
accelerating protons to higher energies than upper point protons.
Therefore, the high energy protons come from  the lower point.
The value of the electric field $E_x$ at the upper position (point-1)
and lower position (point-3) differs by about $30\%$ in the time interval
between 15 and 20.
Figure \ref{fig:J}(b) shows the time change of $\int_0^t E_{x}(\tau)d\tau$
at points 1,2,3
where $E_{x}(\tau)$ has been normalized by the maximum value of $E_x^{fil.}$
in Fig. \ref{fig:J}(a),
indicating how much the accelerating electric field has risen in time.
It is seen that the lower point protons are accelerated by an overall larger
electric field.
In terms of the time integration of $E_x$,
even though the starting point of field is later for
the lower position (point-3) due to the delay of the laser arrival time,
the rate of increase point-3 is higher than the other points.
Point-3 acquires the highest value finally.
The value of the lower position (point-3) is about $2\%$ bigger than
the upper position (point-1) value at time $t=25$.
For $\Delta \mathcal{E}/\mathcal{E}\leq 2\%$
the difference of $\int E_x(t)dt$ should be less than about $1\%$ as mentioned
above. From this considration $2\%$ is a big difference.
Figure \ref{fig:J}(c) shows the value of $E_x^{fil.}$ at the position
$2.5\lambda $ from the target surface (Fig. \ref{fig:I}).
Because these points are some distance from the target surface,
the values of the electric field in Fig. \ref{fig:J}(c)
$E_{x}$ become small compared with those on the target surface.
However, the electric field at the lower position (point-6) is still larger
than at the upper position (point-4).
As shown in Fig. \ref{fig:J}(d) the accelerating electric field
is larger for the protons at the lower points.
However, the electric field $E_x$ and $\int E_x(t)dt$ at this position
are about $1/10$ of that compared with the values on the target surface,
and it can be said that the influences of the differences in the acceleration
fields are smaller than at the target surfaces.

\begin{figure}
\includegraphics[width=8.6cm]{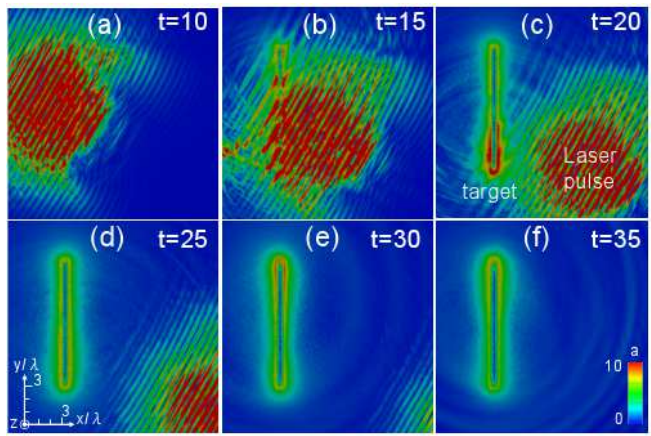}
\caption{\label{fig:L}
Electric field magnitude at early simulation times for case-A.
}
\end{figure}

Figure \ref{fig:K} shows the particle distribution at early times for case-A,
indicating that the electric field in the $x$ direction in the lower
position of the proton layer can be larger. As shown in Fig. \ref{fig:K}(c),
we can see that a lot of electrons which are pushed out from the target are
distributed at the lower position.
This is because the propagation direction of the laser pulse is from the upper
left side to the lower right side, and the electrons pushed out from the target
are distributed more around the lower position.
It is thought that the repulsive force that the protons receive from the gold
target (gold ions) is almost the same at each proton position,
while the force by which the protons are pulled from the electron cloud becomes
larger at lower positions due to this electron distribution.
So, the force that the protons receive in total becomes larger at the lower
part of the proton layer.
Next, after this electron cloud leaves with the laser pulse,
shown in Fig. \ref{fig:K}(f), the number of electrons that exist between the
proton bunch and the gold target is less at the lower position than
the upper position.
This is because the electrons which exist near the center of the laser pulse
are pushed out by the laser pulse, so the number of electrons decreases.
The protons receive a strong repulsion force from the target (gold ions),
and when a lot of electrons are distributed between the proton bunch
and the gold target, this repulsion force is weakened by the electrons.
On the other hand, the repulsion force from the target (gold ions) that
the protons receive becomes strong when the number of electrons between
the proton bunch and the target is small.
Therefore, the protons in the lower part feel a strong electric field,
and are accelerated strongly.

Figure \ref{fig:L} shows the electric field magnitude around the target,
where the times are the same as Fig. \ref{fig:K}.
We can see that the laser pulse progresses from the upper left to the lower
right and the electric field is formed around the target after the laser pulse
passes.
In addition, from Fig. \ref{fig:K} and Fig. \ref{fig:L},
we can see that the electrons in the target are pushed outside
the target by the laser pulse, and a part of the electrons becomes a bunch
and advances with the laser pulse.

\section{CONCLUSIONS}
We have studied the energy spread of protons generated by a laser pulse
obliquely incidence on a double layer target.
We have found that the protons which are initially located at a position
downward in the laser propagation direction from the irradiation laser pulse
center become more energetic.
This is because the accelerating electric field is stronger at the downward
position, owing to the distribution of electrons around the target.
In addition, the maximum energy of the protons is distributed at the position
which is shifted from the laser pulse center position
from the normal to the target surface to the laser propagation direction.
In the oblique incidence case, this is a cause of the energy spread where
the maximum energy of the protons is distributed at a position shifted from
the center line of the target.
In the oblique incidence case, we can reduce the energy spread of the generated
protons by controlling the laser pulse irradiation point.
This provides a way to control the proton energy spread by adjusting the
laser irradiation point.

\section*{Acknowledgements}
The computations were performed with the Power Edge 2950 cluster system
at JAEA Kansai and ALTIX 3700 at JAEA Tokai.

\end{document}